\documentclass[letterpaper, 10 pt, conference]{ieeeconf}  

\IEEEoverridecommandlockouts                              
\usepackage{graphics} 
\usepackage{epsfig} 
\usepackage{mathptmx} 
\usepackage{times} 
\usepackage{amsmath} 
\usepackage{amssymb}  
\usepackage{xcolor}
\usepackage{booktabs}
\usepackage{caption}

\usepackage{algorithm}
\usepackage[noend]{algpseudocode}

\makeatletter
\def\BState{\State\hskip-\ALG@thistlm}
\makeatother

\newcommand*\xbar[1]{%
  \hbox{%
    \vbox{%
      \hrule height 0.5pt 
      \kern0.5ex
      \hbox{%
        \kern-0.1em
        \ensuremath{#1}%
        \kern-0.1em
      }%
    }%
  }%
} 

\DeclareMathOperator*{\argmin}{arg\,min}

\newtheorem{lem}{Lemma}

\title{\LARGE \bf
On-line Bayesian System Identification}

\author{D. Romeres, G. Prando,  G. Pillonetto and A. Chiuso $^\dagger$
\thanks{This work has been partially supported by the FIRB project ``Learning meets time'' (RBFR12M3AC) funded by MIUR.}
\thanks{$^\dagger$ Dept. of Information  Engineering, University of Padova (e-mail: \{\tt \small romeresd,prandogi,giapi,chiuso\}@dei.unipd.it)}%
}

\begin{document}

\maketitle
\thispagestyle{empty}
\pagestyle{empty}

\begin{abstract}
We consider an on-line system identification setting, in which new data become available at given time steps. In order to meet real-time estimation requirements, we propose a tailored Bayesian system identification procedure, in which the hyper-parameters are still updated through Marginal Likelihood maximization, but after only one iteration of a suitable iterative optimization algorithm. Both gradient methods and the EM algorithm are considered for the Marginal Likelihood optimization. We compare this ``1-step'' procedure with the standard one, in which the optimization method is run until convergence to a local minimum. The experiments we perform confirm the effectiveness of the approach we propose.
\end{abstract}

\section{Introduction}

The system identification problem has been addressed for many years by resorting to so-called parametric methods: among them, the most common one is the Prediction Error Method (PEM), where the parameters are estimated by minimizing a functional measuring the prediction errors \cite{Ljung:99,Soderstrom}. Recursive PEM \cite{RecursiveBook} is a well-established variant of the standard PEM approach, which allows to deal with on-line situations, where data are not processed ``in batch'', but  model estimates are computed iteratively as  new data  becomes available. This type of methods can e.g. handle situations in which a sensor provides new measurements at fixed time intervals; another important application of this approach involves the identification of (slowly) time-variant systems, where a real-time tracking of the system dynamics is necessary.\\ It is well known that selecting the model complexity is a critical issue in parametric system identification \cite{Ljung:99,Soderstrom,SS2010,SS2011,ChenOL12,SurveyKBsysid}; the more so in the recursive framework, in particular when the system under analysis is slowly time-varying. In fact  model complexity selection rules, which trade model complexity versus fit, 
may turn out to give different answers as new data becomes available;  of course if the ``true system'' is also time varying one should actually expect that also the estimator follows these variations. Dealing with parametric model classes in which the order changes over time is definitely a delicate (and possibly nontrivial) issue. 
%

Recently, a new non-parametric approach relying on Bayesian estimation techniques has been introduced in the system identification community \cite{SS2011,SurveyKBsysid}. In this work we extend this new framework by introducing an incremental procedure, which is suitable for an on-line setting. In the Bayesian framework hyperparameters, which describe the ``prior'', have first to be estimated in order to compute a posterior distribution of the unknown systems. Of course, if one is interested in a point estimator of the system, then the conditional mean is readily available in closed form in the Gaussian scenario we consider. The hyperparameter estimation, which  replaces the order estimation step in the parametric case, allows to continuously adapt the model complexity as new data become available as well as when the ``true'' underlying system changes over time. 
%

This paper focuses on gradient-based as well as EM-based algorithms for updating the hyperparameter estimates (as well as the system estimate); comparison among these methods will be provided through simulation results
 both in terms of accuracy as well as computational time. Some connections between EM-based, gradient-based methods and iteratively reweighed schemes  will be also provided, showing that there is a strong similarity among these seemingly different approaches.

 The paper is organized as follows. In Section \ref{sec:prob_form} we state the problem we are considering and we briefly review the non-parametric/Bayesian approach for system identification. Section \ref{Sec:onlineSettings} introduces the on-line procedure we will implement, while Section \ref{Sec:1stepMLmax} illustrates the how standard iterative methods are adapted in order to deal with the real-time requirements. In Section \ref{sec:connection} we will outline some connections between gradient methods and the EM algorithm which are typically adopted to solve likelihood optimization problems. Section \ref{sec:results} will present  some experimental results while conclusions and a brief discussion on future research directions are drawn in Section \ref{sec:conclusion}.
 
\section{Problem Formulation}\label{sec:prob_form}
Consider two jointly stationary discrete-time zero-mean stochastic processes $\left\{u(t)\right\}$, $\left\{y(t)\right\}$, $t\in\mathbb{Z}$, and assume that they are respectively the measurable input and output of an Output Error model, i.e.:
\begin{equation}\label{equ:oe_model}
y(t) = \left[h * u\right](t) + e(t), \qquad y(t),u(t)\in\mathbb{R}
\end{equation}
where $h(t)$ is the model impulse response. $e(t)$ is assumed to be a zero-mean Gaussian white noise affecting the output measurements and being uncorrelated to $u(t)$.\\
Standard system identification procedures aim at estimating the impulse response $h(t)$ (or an equivalent representation of the model \eqref{equ:oe_model}) on the basis of a set of input-output data pairs $\left\{u(t),y(t)\right\}_{t=1}^{N}$.\\
In this work we consider the recently introduced non-parametric/Bayesian paradigm for system identification and we adapt it to an on-line identification setting. Namely, assume that at time step $k+1$ a dataset $\mathcal{D}_{k+1}=\left\{u(t),y(t)\right\}_{t=1}^{N_{k+1}}$ becomes available: by means of this new data, we aim at updating the previous system estimate (based on datasets $\left\{\mathcal{D}_i\right\}_{i=1}^k$, while keeping the computational complexity and the memory storage as low as possible.
Next section will briefly introduce the non-parametric/Bayesian approach to system identification \cite{SS2011,SurveyKBsysid}.
\subsection{Bayesian System Identification}
\label{Subsec:bayesianSI}
For simplicity, we approximate the IIR model \eqref{equ:oe_model} with a FIR model of order $n$, thus considering the estimation of $\left\{h(t)\right\}_{t=1}^n$. If $n$ is chosen sufficiently large, the bias arising in the estimate as a consequence of this assumption will be negligible (in particular if the true impulse response $h$ has an exponential decay). The techniques discussed in this section can be extended to the estimation of IIR models by resorting to the theory of Reproducing Kernel Hilbert Spaces (RKHS) \cite{SS2010}.
Under the FIR model assumption we can rewrite the relation between $N_k$ input-output data pairs as a linear regression model, i.e.:
\begin{equation}
\mathbf{y} = \mathbf{\phi} \mathbf{h} + \mathbf{e}
\end{equation}
where 
\begin{align}
\mathbf{y} &:= \left[\begin{array}{ccc} y(1) & \cdots & y(N_k)\end{array}\right]^\top \in\mathbb{R}^{N_k}\label{equ:vectors}\\
\mathbf{h} &:= \left[\begin{array}{ccc} h(1) & \cdots & h(n)\end{array}\right]^\top\mathbb{R}^n\nonumber\\
\mathbf{e} &:= \left[\begin{array}{ccc} e(1) & \cdots & e(N_k)\end{array}\right]^\top \in\mathbb{R}^{N_k}\nonumber\\
\phi &:= \begin{bmatrix}
u(0) & u(-1) & \cdots  & u(-n+1)\\
u(1) & u(0) & \cdots  & u(-n+2)\\
\vdots & \ddots & \ddots  & \vdots\\
u(N_k-1) & u(N_k-2) & \cdots  & u(N_k-n)\\
u(N_k) & u(N_k-1) & \cdots  & u(N_k-n+1)
\end{bmatrix}\in\mathbb{R}^{N_k\times n}\nonumber
\end{align}
Under the Bayesian framework, a prior distribution for the impulse response is first designed in order to account for some desired properties (e.g. smoothness, stability, etc.). A typical choice (inherited from the Gaussian process regression approach) is to postulate a Gaussian distribution:
\begin{align}
p_\eta(\mathbf{h}) &\sim \mathcal{N}(0, K_\eta), \quad  K_\eta=\lambda K_\beta\in\mathbb{R}^{n\times n},\ \lambda \in\mathbb{R},\ \beta\in\mathbb{R}^{d-1}\label{equ:prior}\\
\Omega &= \left\{\eta=[\lambda,\beta]\in\mathbb{R}^{d}: \lambda\geq 0,\ 0\leq \beta \leq 1 \right\}\label{equ:omega}
\end{align}
In \eqref{equ:prior} $\eta$ play the role of hyper-parameters that shape the covariance matrix and need to be estimated using the available data, while $\Omega$ denotes their feasible set. In machine learning literature the covariance matrix $K_\eta$ is typically called kernel.\\
Under the Gaussian assumption for the noise $e(t)$, the joint distribution of $\mathbf{y}$ and $\mathbf{h}$ is jointly Gaussian, for fixed values of $\eta$. This allows to compute the minimum variance estimator of $\mathbf{h}$ in closed form as:
\begin{equation}\label{equ:min_var_est}
\widehat{\mathbf{h}} := \mathbb{E}\left[\mathbf{h}\vert \mathbf{y},\eta\right] = \left( \phi^\top\phi + \sigma^2 K_\eta^{-1}\right)^{-1} \phi^\top \mathbf{y}
\end{equation}
The Bayesian formulation also provides a tool for a robust estimation of the hyper-parameters $\eta$ \cite{PCAuto2015}. This is accomplished by maximizing the so-called marginal likelihood, which is obtained after $\mathbf{h}$ has been integrated out from the joint probability density of $p(\mathbf{y},\mathbf{h})$. Since $\mathbf{y}$ and $h$ are jointly Gaussian, the Marginal Likelihood (ML) is available in closed form, leading to
\hspace{-10mm}\begin{align}
\hat{\eta} &= \arg\max_{\eta\in\Omega} p(\mathbf{y}\vert \eta)\label{equ:ml_max} \\
&\equiv \arg \min_{\eta\in\Omega} - \ln p(\mathbf{y}\vert \eta)= \arg\min_{\eta\in\Omega} \mathbf{y}^\top\Sigma_y(\eta)^{-1} \mathbf{y}+ \ln \det \Sigma_y(\eta) \nonumber\\
&\Sigma_y(\eta)=  \phi K_\eta \phi^\top + \sigma^2 I_{N_k}
\end{align}
Therefore, once the estimate $\hat{\eta}$ in \eqref{equ:ml_max} is computed, it can be plugged in into \eqref{equ:min_var_est} to obtain the so-called Empirical Bayes estimator. Notice that an estimate of the noise variance $\sigma^2$ is also required in order to determine $\widehat{\mathbf{h}}$ in \eqref{equ:min_var_est}. To this purpose one possibility is to treat $\sigma^2$ as an hyper-parameter and to estimate it by means of \eqref{equ:ml_max}; an alternative is to set it as the noise variance estimate computed from a LS estimate of $\mathbf{h}$. In the following we will adopt the latter option. \\
Next section will outline how this estimation framework can be adapted to the on-line system identification setting.

\section{On-Line Setting}
\label{Sec:onlineSettings}
Consider the on-line setting outlined in Section \ref{sec:prob_form}. Assume that a current impulse estimate $\widehat{\mathbf{h}}^{(k)}$ and hyper-parameters estimate $\hat{\eta}^{(k)}$ are available; Algorithm \ref{alg:on_line_setup} summarizes how these estimates can be updated by exploiting the new dataset $\mathcal{D}_{k+1}= \left\{u(t),y(t)\right\}_{t=1}^{N_{k+1}}$. \\
In Algorithm \ref{alg:on_line_setup}, $\phi^{(k)}$ denotes the matrix defined in \eqref{equ:vectors} built with the input data coming from dataset $\mathcal{D}_k$, while we denote with $\Phi^{(k)}\in\mathbb{R}^{\bar{N}_k\times n}$ the matrix built with the inputs coming from the first $k$ datasets $\left\{\mathcal{D}\right\}_{i=1}^k$, with $\xbar{N}_k = \sum_{i=1}^k N_i$. An analogous notation is adopted for $\mathbf{y}^{(k)}$ and $Y^{(k)} \in \mathbb{R}^{\bar{N}_k}$. We also define the useful quantities $R^{(k)}=\Phi^{(k)^\top}\Phi^{(k)}$, $\widetilde{Y}^{(k)}= \Phi^{(k)^\top} Y^{(k)}$, $\xbar{Y}^{(k)} = Y^{(k)^\top }Y^{(k)}$. 
\begin{algorithm}
\caption{On-Line Bayesian System Identification}\label{alg:on_line_setup}
\begin{algorithmic}[1]
\Statex{\textbf{Inputs:}} previous estimates $\left\{ \hat{\eta}^{(k)}, \hat{\eta}^{(k-1)}\right\}$, previous data matrices $\left\{R^{(k)},\widetilde{Y}^{(k)},\xbar{Y}^{(k)}\right\}$, new data $\mathcal{D}_{k+1}=\left\{u(t),y(t);\ t=1,...,N_{k+1}\right\}$
\State $R^{(k+1)} \gets R^{(k)} + \phi^{(k+1)}\phi^{(k+1)^\top}$\label{alg_step:r}
\State $\widetilde{Y}^{(k+1)} \gets \widetilde{Y}^{(k)} +  \phi^{(k+1)}\mathbf{y}^{(k+1)}$\label{alg_step:yt}
\State $\xbar{Y}^{(k+1)} \gets \xbar{Y}^{(k)} + \mathbf{y}^{(k+1)^\top}\mathbf{y}^{(k+1)}$\label{alg_step:yb}
\State $\widehat{\mathbf{h}}_{LS}^{(k+1)} \gets R^{(k+1)^{-1}} \widetilde{Y}^{(k+1)}$ \label{alg_step:ls}
\State {\footnotesize$\hat{\sigma}^{(k+1)^2} \gets \frac{1}{\bar N_k - n} \left(\bar{Y}^{(k+1)}-2\widetilde{Y}^{(k+1)^\top}\widehat{\mathbf{h}}_{LS}^{(k+1)} + \widehat{\mathbf{h}}_{LS}^{(k+1)^\top}R^{(k+1)}\widehat{\mathbf{h}}_{LS}^{(k+1)} \right)$}
\State Compute $\hat{\eta}^{(k+1)}$ through 1-step Marginal Likelihood maximization initialized with $\hat{\eta}^{(k)}$ and $\hat{\eta}^{(k-1)}$\label{alg_step:ml}
\State $\widehat{\mathbf{h}}^{(k+1)} \gets \left(R^{(k+1)} +\hat{\sigma}^{(k+1)^2} K_{\hat{\eta}^{(k+1)}}^{-1}\right)^{-1}\widetilde{Y}^{(k+1)}$\label{alg_step:h}
\Statex{\textbf{Output:}} $\widehat{\mathbf{h}}^{(k+1)}$
\end{algorithmic}
\end{algorithm}

The key step of the procedure outlined in Algorithm \ref{alg:on_line_setup} is the hyper-parameter estimation at step \ref{alg_step:ml}, i.e.
\begin{equation}\label{equ:nml_min}
\hat{\eta}^{(k+1)} = \arg\min_{\eta\in\Omega} L(\eta) = \arg\min_{\eta\in\Omega} -\ln p(Y^{(k+1)}\vert \eta)
\end{equation}
Typically, the ML maximization required at that step is performed by adopting iterative methods, such as 1st or 2nd order optimization algorithms or the Expectation-Maximization (EM) algorithm. However, both these approaches could require a large number of iterations before reaching convergence, thus significantly increasing the computational complexity of Algorithm \ref{alg:on_line_setup}. Recall that the ML can be robustly evaluated with computational complexity $O(n^3)$ as \cite{chen2013implementation}
\begin{align*}
L(\eta) &= (\xbar{N}_{k+1}-m) \ln \hat{\sigma}^{(k+1)^2} + 2\ln \det S +\\
&+\frac{1}{\hat{\sigma}^{(k+1)^2}}\left( \xbar{Y}^{(k+1)} - \widetilde{Y}^{(k+1)^\top} LS^{-\top} S^{-1}L^\top\widetilde{Y}^{(k+1)}\right)
\end{align*}
where $L$ and $S$ are Cholesky factors:
$$K_\eta=LL^\top, \quad\hat{\sigma}^{(k+1)^2}I_n + L^\top R^{(k+1)}L=SS^\top$$
Therefore, if an optimization algorithm is adopted for ML maximization, each iteration would have complexity $O(n^3)$.\\
To accelerate hyper-parameters estimation, at step \ref{alg_step:ml} of Algorithm \ref{alg:on_line_setup} we just perform one iteration of these iterative methods. In particular, we will compare the performances of some 1st order methods and of the EM algorithm. Next section will illustrate them in more details.

For what regards the computational complexity of the remaining steps in Algorithm \ref{alg:on_line_setup}, the  most demanding ones are steps \ref{alg_step:ls} and \ref{alg_step:h}, which are both $O(n^3)$, because of the matrix inversion that has to be computed. If the new dataset $\mathcal{D}_{k+1}$ consists on only one input-output pair, then Shermann-Morrison formula can be exploited to compute $R^{(k+1)^{-1}}$ with a complexity of $O(n^2)$. \\
Furthermore, notice that the memory storage requirements of Algorithm \ref{alg:on_line_setup} are $O(n^2)$, thanks to the updates at steps \ref{alg_step:r}-\ref{alg_step:yb}.

\section{1-step Marginal Likelihood Maximization}
\label{Sec:1stepMLmax}
In this work we consider two different approaches to solve problem \eqref{equ:ml_max}: 1st order optimization algorithms (also known as gradient methods) and the EM algorithm, which is suited to compute maximum likelihood solutions for models having latent variables. As previously anticipated in the previous section, we will only perform one iteration of these algorithms, in order to address the on-line requirement that our setting imposes. 
The two approaches are now described.

\subsection{Gradient Methods}
The one-step implementation of a gradient method we consider is summarized in Algorithm \ref{alg:1step_grad}. We should stress the fact that, in our setting, apex $^{(k)}$ refers to the value taken by a certain quantity after $k$ datasets $\left\{\mathcal{D}_i\right\}_{i=1}^k$ have been seen; it does not refer to the iteration number of the considered gradient method (since we are performing just one iteration).\\
Notice that the update rule we use in Algorithm \ref{alg:1step_grad} for $\hat{\eta}^{(k)}$ is a Quasi-Newton method: specifically, at step \ref{alg_step:inverse_H} we just compute an approximation to the inverse Hessian, unlike Newton's update rule which requires the exact Hessian computation.\\
Quasi-Newton methods approximate the Hessian by using only gradient information. Different algorithms can be derived according to the specific Hessian approximation that is chosen. They essentially differ in the way in which they attempt to satisfy the so-called \textit{secant equation} \cite{Nocedal06}:
\begin{equation}\label{equ:secant_eq}
B^{(k)}w^{(k-1)} = r^{(k-1)}
\end{equation}
where $B^{(k)}$ represents the approximation to the inverse Hessian computed at $\hat{\eta}^{(k)}$, while
$$
r^{(k-1)} = \hat{\eta}^{(k)} - \hat{\eta}^{(k-1)}, \qquad
w^{(k-1)} = \nabla L( \hat{\eta}^{(k)}) - \nabla L( \hat{\eta}^{(k-1)})
$$
In the following we will illustrate the three different procedures we consider to approximate the inverse Hessian. According to the chosen approximation, the projection operator $\Pi_{\Omega,W}$ onto the feasible set $\Omega$ at step \ref{alg_step:proj} changes; namely, it is defined as:
\begin{equation}\label{equ:proj}
\Pi_{\Omega,W}(z) = \argmin_{x\in\Omega} (x-z)^\top W(x-z)
\end{equation}
and the matrix $W$ takes different values according to how $B^{(k)}$ is computed.

\begin{algorithm}
\caption{1-step Gradient Method}\label{alg:1step_grad}
\begin{algorithmic}[1]
\Statex{\textbf{Inputs:}} previous estimates $\left\{ \hat{\eta}^{(k)}, \hat{\eta}^{(k-1)}\right\}$, $\nabla L(\hat{\eta}^{(k-1)})$, $L(\hat{\eta}^{(k-1)})$, $R^{(k+1)}$, $\widetilde{Y}^{(k+1)}$, $\xbar{Y}^{(k+1)}$, $\hat{\sigma}^{(k+1)^2}$
\Statex Initialize parameters $c$ and $\delta$
\State Compute $\nabla L(\hat{\eta}^{(k)})$
\State $r^{(k-1)} \gets \hat{\eta}^{(k)} - \hat{\eta}^{(k-1)}$ \label{alg_step:rf}
\State $w^{(k-1)} \gets \nabla L(\hat{\eta}^{(k)}) - \nabla L(\hat{\eta}^{(k-1)})$ \label{alg_step:w}
\State Compute the inverse Hessian approximation $B^{(k)}$ using one among Algorithm \ref{alg:bb},\ref{alg:sgp},\ref{alg:bfgs}
\label{alg_step:inverse_H}
\State Project onto the feasible set:
\Statex $z\gets \Pi_{\Omega,W} \left(\hat{\eta}^{(k)} -\nabla L(\hat{\eta}^{(k)})\right)$\label{alg_step:proj}
\State $\Delta\hat{\eta}^{(k)} \gets z - \hat{\eta}^{(k)}$
\State $\gamma \gets 1$
\If{$L(\hat{\eta}^{(k)}+\gamma \Delta \hat{\eta}^{(k)}) \leq L(\hat{\eta}^{(k)})+ c \gamma \nabla(\hat{\eta}^{(k)})^\top \Delta\hat{\eta}^{(k)}$}
\State Go to step 12
\Else
\State $\gamma \gets \delta \gamma$
\EndIf
\State $\hat{\eta}^{(k+1)} \gets \hat{\eta}^{(k)} + \gamma \Delta \hat{\eta}^{(k)}$
\Statex \textbf{Output:} $\hat{\eta}^{(k+1)}$
\end{algorithmic}
\end{algorithm}

\subsubsection{Barzilai-Borwein (BB) \cite{barzilai1988two}} This approach approximates the inverse Hessian by simply computing an appropriate step-size $\alpha^{(k)}>0$, i.e. $B^{(k)}=\alpha^{(k)} I_d$ and $\alpha^{(k)}$ is set to be the solution of one of the following two problems:
\begin{align}
\alpha_1^{(k)} &:= \argmin_\alpha \|\alpha r^{(k-1)} - w^{(k-1)}\|^2 = \frac{r^{(k-1)^\top}r^{(k-1)}}{r^{(k-1)^\top}w^{(k-1)}}\\
\alpha_2^{(k)} &:= \argmin_\alpha \|r^{(k-1)} - \alpha w^{(k-1)}\|^2 = \frac{r^{(k-1)^\top}w^{(k-1)}}{w^{(k-1)^\top}w^{(k-1)}}
\end{align}
Our implementation (outlined in Algorithm \ref{alg:bb}) follows the alternation strategy proposed in \cite{BonettiniCPSIAM2014}, where both $\alpha_1$ and $\alpha_2$ are alternatively chosen. 
In this case, the matrix $W$ in the projection $\Pi_{\Omega,W}$ \eqref{equ:proj} is set equal to the identity matrix $I_d$.

\begin{algorithm}
\caption{Barzilai-Borwein Alternation Strategy}\label{alg:bb}
\begin{algorithmic}[1]
\Statex \textbf{Inputs:} $\tau^{(k)}, \hat{r}^{(k-1)}, \hat{w}^{(k-1)}$
\Statex Set $0<\alpha_{min}<\alpha_{max}$
\State $\alpha_1 \gets \left(r^{(k-1)^\top}r^{(k-1)}\right)/\left(r^{(k-1)^\top}w^{(k-1)}\right)$
\State $\alpha_2 \gets \left(r^{(k-1)^\top}r^{(k-1)}\right)/\left(w^{(k-1)^\top}w^{(k-1)}\right)$
\State $\tilde{\alpha}_1 \gets \min \left\{ \max\left\{\alpha_{min},\alpha_1\right\},\alpha_{max}\right\}$
\State $\tilde{\alpha}_2 \gets \min \left\{ \max\left\{\alpha_{min},\alpha_2\right\},\alpha_{max}\right\}$
\If{$\tilde{\alpha}_2/\tilde{\alpha}_1\leq \tau^{(k)}$}
\State $\alpha^{(k)} \gets \tilde{\alpha}_2$
\State $\tau^{(k+1)} \gets 0.9 \tau^{(k)}$
\Else
\State $\alpha^{(k)} \gets \tilde{\alpha}_1$
\State $\tau^{(k+1)} \gets 1.1 \tau^{(k)}$
\EndIf
\Statex \textbf{Outputs:} $B^{(k)}=\alpha^{(k)} I_d$, $\tau^{(k+1)}$
\end{algorithmic}
\end{algorithm}

\subsubsection{Scaled Gradient Projection (SGP) \cite{BonettiniCPSIAM2014}}
When adopting the Scaled Gradient Projection method, the inverse Hessian approximation $B^{(k)}$ at step \ref{alg_step:inverse_H} of Algorithm \ref{alg:1step_grad} is computed as:
\begin{equation}
B^{(k)} = \alpha^{(k)} D^{(k)}, \qquad \alpha^{(k)}\in\mathbb{R}^+, \ \ D^{(k)}\in\mathbb{R}^{d\times d}
\end{equation}
The step-size $\alpha^{(k)}$ is again computed by using the alternated Barzilai-Borwein rules above illustrated. The exact implementation is slightly different from the ones outlined in Algorithm \ref{alg:bb}, due to the presence of the matrix $D^{(k)}$ (refer to \cite{BonettiniCPSIAM2014} for the exact implementation). $D^{(k)}$ is a scaling matrix whose choice strictly depends on the objective function and on the constraints set of the problem we are considering. Our implementation follows the one proposed in \cite{BonettiniCPSIAM2014}, where $D^{(k)}$ is a diagonal matrix with the diagonal entries chosen according to the gradient split idea.\\
Consider the problem \eqref{equ:nml_min} and let us define $D^{(k)} = blockdiag(D_\lambda^{(k)},D_\beta^{(k)})$ where $D_\lambda^{(k)}\in\mathbb{R}$ and $D_\beta^{(k)}\in\mathbb{R}^{(d-1)\times (d-1)}$ respectively denote the scaling matrices built for the two components of the hyper-parameter vector $\eta$.
In the following we will briefly outline the definition of matrix $D_\lambda^{(k)}$ in relation to the non-negative constraint $\lambda\geq 0$. Refer to \cite{BonettiniCPSIAM2014} for the derivation of $D_\beta^{(k)}$, since the box constraints in \eqref{equ:omega} have to be considered.\\
The definition of $D_\lambda^{(k)}$ relies on the following decomposition of the gradient w.r.t. $\lambda$ of the objective function $L(\eta)$ in \eqref{equ:nml_min}:
\begin{align}
\nabla_\lambda L(\eta) &= V_\lambda(\eta) - U_\lambda(\eta)\label{equ:grad_decomp}\\
V_\lambda(\eta) &= \nabla_\lambda \left(\ln \det \Sigma_y(\eta)\right)\label{equ:v_lambda}\\
&=\mbox{Tr} \left(\Sigma_y(\eta)^{-1} \Phi^{(k+1)}K_\beta \Phi^{(k+1)^\top}\right) >0\nonumber\\
U_\lambda(\eta) &= - \nabla_\lambda \left(Y^{(k+1)^\top} \Sigma_y(\eta) Y^{(k+1)}\right)\label{equ:u_lambda}\\
&= Y^{(k+1)^\top}\Sigma_y(\eta)^{-1} \Phi^{(k+1)}K_\beta \Phi^{(k+1)^\top} \Sigma_y(\eta)^{-1} Y^{(k+1)}\geq 0  \nonumber
\end{align}
where $\nabla_\lambda$ denotes the gradient w.r.t. $\lambda$. Notice that the above inequalities hold because of the positive semi-definiteness of $K_\beta$.\\
In view of decomposition \eqref{equ:grad_decomp}, the first order optimality conditions w.r.t. $\lambda$ for problem \eqref{equ:nml_min}, i.e.
\begin{equation}
\lambda \nabla_\lambda L(\eta) = 0, \quad \lambda\geq 0, \quad \nabla_\lambda L(\eta)\geq 0
\end{equation}
can be rewritten as the fixed point equation $\lambda=\lambda U_\lambda(\eta)/V_\lambda(\eta)$. By exploiting the fixed point update method, we can then define
\begin{equation}
D_\lambda^{(k)} = \min\left\{\max\left\{d_{min},\frac{\hat{\lambda}^{(k)}}{V_\lambda(\hat{\eta}^{(k)})}\right\},d_{max}\right\}
\end{equation}
Refer to \cite{BonettiniCPSIAM2014} for a more detailed derivation.\\
Algorithm \ref{alg:sgp} summarizes how $B^{(k)}$ at step \ref{alg_step:inverse_H} of Algorithm \ref{alg:1step_grad} is computed through SGP. In this case $\Pi_{\Omega,W}$ at step \ref{alg_step:proj} is defined setting $W=D^{(k)^{-1}}$.

\begin{algorithm}
\caption{Scaled Gradient Projection Algorithm (SGP)}\label{alg:sgp}
\begin{algorithmic}[1]
\Statex \textbf{Inputs:} $\nabla L(\hat{\eta}^{(k)}), \ \tau^{(k)}, \hat{r}^{(k-1)}, \hat{w}^{(k-1)}$
\Statex Set $0<d_{min}<d_{max}$
\State Compute $V_\lambda(\hat{\eta}^{(k)})$ as in \eqref{equ:v_lambda}
\State Compute $U_\lambda(\hat{\eta}^{(k)})$ as in \eqref{equ:u_lambda}
\State $D_\lambda^{(k)} \gets \min\left\{\max\left\{d_{min},\frac{\hat{\lambda}^{(k)}}{V_\lambda(\hat{\eta}^{(k)})}\right\},d_{max}\right\}$
\State Compute $V_\beta(\hat{\eta}^{(k)})>0$ and $U_\beta(\hat{\eta}^{(k)})>0$ s.t. $\nabla_\beta L(\hat{\eta}^{(k)})= V_\beta(\hat{\eta}^{(k)}) - U_\beta(\hat{\eta}^{(k)})$
\State Compute $D_\beta^{(k)}$ as illustrated in \cite{BonettiniCPSIAM2014}
\State $D^{(k)} \gets blockdiag(D_\lambda^{(k)},D_\beta^{(k)})$
\State Run Algorithm \ref{alg:bb} to compute $\alpha^{(k)},\ \tau^{(k+1)}$
\Statex \textbf{Outputs:} $B^{(k)}=\alpha^{(k)} D^{(k)}$, $\tau^{(k+1)}$
\end{algorithmic}
\end{algorithm}
\vspace{-2mm}
\subsubsection{BFGS}
When adopting the inverse Hessian approximation provided by BFGS method, $B^{(k)}$ at step \ref{alg_step:inverse_H} of Algorithm \ref{alg:1step_grad} is computed as the unique solution of the following problem
\begin{align} 
&\min_B \|B-B^{(k-1)}\|_{\mathcal{M}} \label{equ:bfgs_probl}\\
&s.t. \ \ B=B^\top,\ B\succ 0,\ B w^{(k-1)}=r^{(k-1)}  \nonumber
\end{align}
where $\|A \|_{\mathcal{M}}=\|\mathcal{M}^{1/2}A\mathcal{M}^{1/2}\|_F$ denotes the weighted Frobenius norm, with $W$ chosen such that $\mathcal{M}r^{(k-1)}=w^{(k-1)}$ \cite{Nocedal06}.
Algorithm \ref{alg:bfgs} summarizes the implementation of BFGS. The projection operator $\Pi_{\Omega,W}$ is in this case defined with $W=I_d$.

\begin{algorithm}
\caption{BFGS}\label{alg:bfgs}
\begin{algorithmic}[1]
\Statex \textbf{Inputs:} $B^{(k-1)},\ \hat{r}^{(k-1)},\ \hat{w}^{(k-1)}$
\State $\rho \gets 1/(\hat{w}^{(k-1)^\top}\hat{r}^{(k-1)})$
\State $B^{(k)} \gets  \rho \hat{r}^{(k-1)}\hat{r}^{(k-1)^\top}+\left(I-\rho \hat{r}^{(k-1)} \hat{w}^{(k-1)^\top}\right) B^{(k-1)} \left(I-\rho \hat{w}^{(k-1)} \hat{r}^{(k-1)^\top}\right)$
\Statex \textbf{Outputs:} $B^{(k)}$
\end{algorithmic}
\end{algorithm}
\vspace{-3mm}

\subsection{EM Algorithm}
The Expectation-Maximization (EM) algorithm is used to compute maximum likelihood solutions for models having latent variables. Recall that at step \ref{alg_step:ml} of Algorithm \ref{alg:on_line_setup}
we need to compute $\hat{\eta}^{(k+1)}$ by maximizing
\begin{align}
p(Y^{(k+1)}\vert \eta) &= \mathbb{E}_{p(\mathbf{h}\vert\eta)} p(Y^{(k+1)},\mathbf{h}\vert\eta)\\
&=\int  p(Y^{(k+1)},\mathbf{h}\vert\eta) p(\mathbf{h}\vert\eta)d\mathbf{h}\nonumber
\end{align}
where we used the notation $\mathbb{E}_q$ to indicate the expectation w.r.t. the probability distribution $q$.
Hence, in our setting $\mathbf{h}$ plays the role of the latent variable. Consider the following decomposition \cite{bishop2006PRML}:
\begin{align}
\ln p(Y^{(k+1)}\vert \eta)&=\mathcal{L}(q(\mathbf{h}),\eta) + KL (q(\mathbf{h})||p(\mathbf{h}\vert Y^{(k+1)},\eta))\nonumber\\
&=\int q(\mathbf{h})\ln \left\{\frac{p(Y^{(k+1)},\mathbf{h}\vert \eta)}{q(\mathbf{h})}\right\}d\mathbf{h}\nonumber\\
&- \int q(\mathbf{h}) \ln \left\{\frac{p(\mathbf{h}\vert Y^{(k+1)},\eta)}{q(\mathbf{h})}\right\} d\mathbf{h}
\end{align}
where $\mathcal{L}(q,\eta)$ represents a lower bound for $\ln p(Y^{(k+1)}\vert \eta)$, while $KL(\cdot||\cdot)$ denotes the Kullback-Leibler divergence between two probability distributions.\\
A standard EM algorithm finds the optimal value for $\eta$ by keeping alternating two steps, namely the Expectation (E) and the Maximization (M) steps, until convergence is reached.
According to our ``1-step'' approach, when we adopt EM at step \ref{alg_step:inverse_H} of Algorithm \ref{Sec:1stepMLmax}, we just perform one E-step and one M-step. Specifically, in the E-step we compute
\begin{align}
\mathcal{L}&\left(p(\mathbf{h}\vert Y^{(k+1)},\hat{\eta}^{(k)}),\eta\right) = \label{equ:em_estep1}\\
&=\mathbb{E}_{p(\mathbf{h}\vert Y^{(k+1)},\hat{\eta}^{(k)})} \left[ \ln p(Y^{(k+1)}\vert \mathbf{h},\eta) + \ln p(\mathbf{h}\vert \eta)\right]\nonumber\\
&- \mathbb{E}_{p(\mathbf{h}\vert Y^{(k+1)},\hat{\eta}^{(k)})} \left[\ln p(\mathbf{h}\vert Y^{(k+1)} ,\hat{\eta}^{(k)})\right]\nonumber
\end{align}
Recalling that $p(Y^{(k+1)}\vert \mathbf{h},\eta) \sim \mathcal{N} (\Phi^{(k+1)}\mathbf{h},\sigma^2 I_{\xbar{N}_{k+1}} )$, using the prior $p(\mathbf{h}\vert\eta)$ in \eqref{equ:prior} and assuming a non-informative prior on $\eta$, we have
\begin{align}
\mathcal{L}&\left(p(\mathbf{h}\vert Y^{(k+1)},\hat{\eta}^{(k)}),\eta\right) = \label{equ:em_estep2}\\
&= -\frac{\xbar{N}_{k+1}}{2}\ln\sigma^2 -\frac{1}{2\sigma^2}\|Y^{(k+1)}\|^2 + \frac{1}{\sigma^2}Y^{(k+1)^\top}\Phi^{(k+1)} \widehat{\mathbf{h}}^{(k)} \nonumber\\
&-\frac{1}{2\sigma^2} \left(\mbox{tr}\left\{\Phi^{(k+1)^\top}\Phi^{(k+1)} P^{(k)}\right\} + \widehat{\mathbf{h}}^{(k)^\top}\Phi^{(k+1)^\top} \Phi^{(k+1)} \widehat{\mathbf{h}}^{(k)^\top}\right)\nonumber\\
& -\frac{1}{2}\ln \det K_\eta -\frac{1}{2} \left(\mbox{tr}\left\{K_\eta^{-1}  P^{(k)}\right\} + \widehat{\mathbf{h}}^{(k)^\top}K_\eta^{-1} \widehat{\mathbf{h}}^{(k)^\top}\right)\nonumber\\
& +\frac{1}{2}\ln \det P^{(k)} +\frac{n}{2} \nonumber
\end{align}
where we have used $P^{(k)}=\left(\sigma^{-2} \Phi^{(k+1)^\top}\Phi^{(k+1)}+K_{\hat{\eta}^{(k)}}\right)^{-1}$.\\
Notice now that this step corresponds to solve
\begin{equation}\label{equ:em_estep3}
\mathcal{L}\left(p(\mathbf{h}\vert Y^{(k+1)},\hat{\eta}^{(k)}),\eta\right) = \max_{q(\mathbf{h})}\mathcal{L}(q(\mathbf{h}),\hat{\eta}^{(k)})
\end{equation}
since $ KL (q(\mathbf{h})||p(\mathbf{h}\vert Y^{(k+1)},\eta))=0$ when $q(\mathbf{h})$ is the posterior distribution obtained with $\hat{\eta}^{(k)}$.\\
In the M-step of the EM algorithm we instead update the hyper-parameters value:
\begin{equation}\label{equ:em_mstep}
\hat{\eta}^{(k+1)} = \arg\max_{\eta\in\Omega} \mathcal{L}(p(\mathbf{h}\vert Y^{(k+1)},\hat{\eta}^{(k)}), \eta)
\end{equation}
The 1-step EM algorithm we adopt to perform step \ref{alg_step:ml} of Algorithm \ref{alg:on_line_setup} is summarized in Algorithm \ref{alg:em}. In our implementation we replace $\sigma^2$ with $\hat{\sigma}^{(k+1)^2}$.

\begin{algorithm}
\caption{BFGS}\label{alg:em}
\begin{algorithmic}[1]
\Statex \textbf{Inputs:} $\hat{\eta}^{(k)}$, $R^{(k+1)}$, $\widetilde{Y}^{(k+1)}$, $\xbar{Y}^{(k+1)}$, $\hat{\sigma}^{(k+1)^2}$
\State \textbf{E-step:} Compute $\mathcal{L}(\left(p(\mathbf{h}\vert Y^{(k+1)},\hat{\eta}^{(k)}),\eta\right)$ as in \eqref{equ:em_estep2}
\State \textbf{M-step:} $\hat{\eta}^{(k+1)} \gets \arg\max_{\eta\in\Omega} \mathcal{L}(p(\mathbf{h}\vert Y^{(k+1)},\hat{\eta}^{(k)}), \eta)$
\Statex \textbf{Outputs:} $\hat{\eta}^{(k+1)}$
\end{algorithmic}
\end{algorithm}
\vspace{-3mm}

\section{Connections with existing methodologies}\label{sec:connection}
In this section we assume to fix the hyper-parameter $\beta$ in \eqref{equ:prior} (its value will be denoted with $\hat{\beta}$) and we only consider the update of the scaling factor $\lambda$. Under this assumption we show how the EM update rule coincides with a gradient-based update if a specific step-size $\alpha^{(k)}$ is chosen. In addition we point out a connection between the EM algorithm and the iterative reweighted methods, which have been introduced for compressive sensing applications \cite{candes2008enhancing,chartrand2008iteratively}.

\subsection{Connection between EM and Gradient Methods}
Consider the EM update rule in \eqref{equ:em_mstep} and assume $K_\eta = \lambda K_{\hat{\beta}}$ (i.e. $\beta$ is fixed). Then the optimization problem \eqref{equ:em_mstep} can be reformulated as
\begin{align}
\hat{\lambda}_{EM}^{(k+1)} = \arg&\max_{\lambda\geq 0}\ -\ln \det(\lambda K_{\hat{\beta}})\\
& -\left(\mbox{tr}\left\{(\lambda K_{\hat{\beta}})^{-1}  P^{(k)}\right\} + \widehat{\mathbf{h}}^{(k)^\top}(\lambda K_{\hat{\beta}})^{-1} \widehat{\mathbf{h}}^{(k)}\right)\nonumber
\end{align}
from which
\begin{equation}\label{equ:em_lambda_update}
\hat{\lambda}_{EM}^{(k+1)} =\frac{1}{n}\left[\widehat{\mathbf{h}}^{(k)^\top} K_{\hat{\beta}}^{-1} \widehat{\mathbf{h}}^{(k)} + \mbox{tr}\left\{(K_{\hat{\beta}})^{-1}  P^{(k)}\right\}\right]
\end{equation}
Notice that the first term in the update rule \eqref{equ:em_lambda_update} corresponds to the current approximation of the value of $\lambda$ which asymptotically maximizes the Marginal Likelihood, i.e. $\hat{\lambda}^* = \frac{1}{n} \mathbf{h}^\top K_{\hat{\beta}}^{-1} \mathbf{h}$, with $\mathbf{h}$ denoting the true impulse response \cite{Aravkin12}. The second term in \eqref{equ:em_lambda_update} instead accounts for the uncertainty in the $\lambda$ estimate, due to the use of a finite amount of data.\\
Consider now the gradient update rule for $\hat{\lambda}^{(k+1)}$ (based on the minimization of the function $L(\lambda)$ defined in \eqref{equ:nml_min}):
\begin{equation}\label{equ:grad_update_lambda}
\hat{\lambda}_{GR}^{(k+1)} = \hat{\lambda}^{(k)} - \alpha_\lambda^{(k)} \nabla_\lambda L(\hat{\lambda}^{(k)})
\end{equation}
We have the following result.

\begin{lem} If $\alpha_\lambda^{(k)}=\frac{(\hat{\lambda}^{(k)})^2}{n}$ in \eqref{equ:grad_update_lambda}, then $\hat{\lambda}_{GR}^{(k+1)}=\hat{\lambda}_{EM}^{(k+1)}$.
\end{lem}

\textit{Proof:} From \eqref{equ:grad_decomp}-\eqref{equ:u_lambda}, letting $\eta=\lambda$ and fixing $\beta$ to $\hat{\beta}$, we have:
\begin{align*}
\nabla_\lambda &L(\hat{\lambda}^{(k)}) = \frac{n}{\hat{\lambda}^{(k)}} -\frac{1}{(\hat{\lambda}^{(k)})^2}\mbox{Tr}\left\{K_{\hat{\beta}}^{-1}P^{(k)}\right\}\\
&-\frac{1}{(\hat{\lambda}^{(k)})^2} \frac{Y^{(k+1)^\top} \Phi^{(k+	1)} P^{(k)}}{\sigma^2} K_{\hat{\beta}}^{-1}  \frac{P^{(k)}\Phi^{(k+1)^\top} Y^{(k+1)}}{\sigma^2}
\end{align*}
Now, introducing this value into \eqref{equ:grad_update_lambda} gives the result.

\subsection{\hspace{-1.8mm}Connection between EM and Iterative Reweighted Methods}
Iterative reweighted methods have been quite recently introduced in the compressive sensing field in order to improve the recovery of sparse solutions. Here we focues on the $\ell_2$-reweighted scheme that has been proposed in \cite{WipfN10} for Sparse Bayesian Learning (SBL) \cite{Tipping01}.
Consider the optimization problem \eqref{equ:nml_min}; since in the current setting $\beta$ is fixed, we have:
$$
\min_{\lambda\geq 0} -\ln p(Y^{(k+1)}\vert \lambda)= \min_{\lambda \geq 0} Y^{(k+1)^\top} \Sigma_y(\lambda)^{-1} Y^{(k+1)} + \ln\det \Sigma_y(\lambda)
$$
Notice that (\cite{Tipping01}, Appendix A)
\begin{align*}
Y^{(k+1)^\top} \Sigma_y(\lambda)^{-1} Y^{(k+1)} = \min_{\mathbf{h}} \frac{1}{\sigma^2}&\|Y^{(k+1)}-\Phi^{(k+1)}\mathbf{h}\|_2^2 +\\
&+\mathbf{h}^{\top}(\lambda K_{\hat{\beta}})^{-1} \mathbf{h}
\end{align*}
Thus, we have
\begin{align*}
\min_{\lambda\geq 0} -\ln p(Y^{(k+1)}\vert \lambda)  
&= \min_{\lambda\geq 0,\mathbf{h}}& \frac{1}{\sigma^2}&\|Y^{(k+1)}-\Phi^{(k+1)} \mathbf{h}\|_2^2 +\\
&&+& \mathbf{h}^\top (\lambda K_{\hat{\beta}})^{-1}\mathbf{h} + \ln \det \Sigma_y(\lambda)\\
&= \min_{\mathbf{h}} &\frac{1}{\sigma^2} &\|Y^{(k+1)}-\Phi^{(k+1)} \mathbf{h}\|_2^2 + g(\mathbf{h})
\end{align*}
where $g(\mathbf{h}) = \min_{\lambda\geq 0}\mathbf{h}^\top (\lambda K_{\hat{\beta}})^{-1}\mathbf{h} + \ln \det \Sigma_y(\lambda)$ is a non-separable penalty function, since it can not be expressed as a summation over functions of the individual impulse response coefficients $\mathbf{h}_i$. Furthermore, it is a non-decreasing concave function of $\mathbf{h}^2 := [h(1)^2\ \cdots \ h(n)^2]^\top$, thus allowing to employ iterative reweighted $\ell_2$ schemes to minimize the function above. Namely,
\begin{align}
g(\mathbf{h})&\leq \mathbf{h}^\top (\lambda K_{\hat{\beta}})^{-1}\mathbf{h} + \ln \det \Sigma_y(\lambda)\nonumber\\
&= \mathbf{h}^\top (\lambda K_{\hat{\beta}})^{-1}\mathbf{h} + \ln \det (\lambda K_{\hat{\beta}}) + \label{equ:silvester} \\
&+\ln\det \left( \frac{\Phi^{(k+1)^\top} \Phi^{(k+1)}}{\sigma^2} + (\lambda K_{\hat{\beta}})^{-1}\right) + \xbar{N}_{k+1}\ln \sigma^2\nonumber\\
&\leq \mathbf{h}^\top (\lambda K_{\hat{\beta}})^{-1}\mathbf{h} + \ln \det (\lambda K_{\hat{\beta}}) + z\lambda^{-1} - s^*(z) + \xbar{N}_{k+1}\ln \sigma^2 \label{equ:bound}
\end{align}
where $s^*(z)$ denotes the concave conjugate of $s(a) := \ln\det \left( \frac{\Phi^{(k+1)^\top} \Phi^{(k+1)}}{\sigma^2} + a K_{\hat{\beta}}^{-1}\right) $, $a=\lambda^{-1}$, given by:
$$s^*(z) = \min_a za -  \ln\det \left( \frac{\Phi^{(k+1)^\top} \Phi^{(k+1)}}{\sigma^2} + a K_{\hat{\beta}}^{-1}\right), \ a=\lambda^{-1}$$
Notice that in \eqref{equ:silvester} the Silvester's determinant identity is used and the bound \eqref{equ:bound} holds for all $z,\lambda \geq 0$.
Hence, we have
\begin{align}
\min_{\lambda\geq 0} -\ln p&(Y^{(k+1)}\vert \lambda)  
= \min_{\lambda\geq 0,z\geq 0,\mathbf{h}}\ \frac{1}{\sigma^2}\|Y^{(k+1)}-\Phi^{(k+1)} \mathbf{h}\|_2^2 +\nonumber\\
&\mathbf{h}^\top (\lambda K_{\hat{\beta}})^{-1}\mathbf{h} + \ln \det (\lambda K_{\hat{\beta}}) + z\lambda^{-1} - s^*(z) \label{equ:reweighting}
\end{align}
where we have omitted the terms that are not relevant to the optimization problem.
We can now state the analogies with the two steps of the EM algorithm. Specifically, recall that the E-step in the EM is equivalent to solving problem \eqref{equ:em_estep3}: the solution is given by the posterior distribution of $\mathbf{h}$ given $\hat{\lambda}^{(k)}$, i.e. $p(\mathbf{h}\vert Y^{(k+1)},\hat{\lambda}^{(k)})$. Analogously, solving \eqref{equ:reweighting} w.r.t. $\mathbf{h}$ for fixed $\hat{\lambda}^{(k)}$ leads to an a-posteriori estimate, namely the Empirical Bayes estimator $\widehat{\mathbf{h}}^{(k+1)}=\mathbb{E}[\mathbf{h}\vert Y^{(k+1)},\hat{\lambda}^{(k)}]$, which coincides with the Maximum a Posteriori estimator of $\mathbf{h}$.\\
On the other hand, solving \eqref{equ:reweighting} for fixed $\widehat{\mathbf{h}}^{(k)}$ leads to
\begin{equation}\label{equ:lambda_update_reweighted}
\hat{\lambda}^{(k+1)} = \frac{1}{n}\left(\widehat{\mathbf{h}}^{(k)^\top}K_{\hat{\beta}}^{-1}\widehat{\mathbf{h}}^{(k)} + z^*\right)
\end{equation}
where \cite{WipfN10}
$$
z^* = \frac{\partial}{\partial a}\ln \det \left(  \frac{\Phi^{(k+1)^\top} \Phi^{(k+1)}}{\sigma^2} + a K_{\hat{\beta}}^{-1}\right) = \mbox{Tr}\left\{P^{(k)}K_{\hat{\beta}}^{-1}\right\}
$$
Thus, the update \eqref{equ:lambda_update_reweighted} coincides with the M-step in \eqref{equ:em_mstep}.

\section{Experimental Results}\label{sec:results}
In this section we report the results obtained by Bayesian procedures in the on-line setting illustrated in Algorithm \ref{alg:on_line_setup}. Specifically, we compare the procedure which estimates the hyper-parameters by means of a standard iterative algorithm (such as SGP, BB, BFGS and EM) and the one which instead performs only one iteration of the above-mentioned methods (such as illustrated in Algorithms \ref{alg:1step_grad} and \ref{alg:em}). In the following we will refer to the first procedure as OPT, while we will use the notation 1-STEP to refer to the latter one.\\
In all the simulations that follow the OPT procedure exploits the SGP algorithm to maximize the Marginal Likelihood.

 %
%
In our experiments we adopt a zero-mean Gaussian prior with a covariance matrix given by the so-called TC-kernel \cite{ChenOL12}:
\begin{equation}
\bar K_\eta^{TC}(k,j) = \lambda \min(\beta^{j},\beta^{j})
\end{equation}
where $\lambda \geq 0$ and $0 \leq \beta \leq 1$ are the hyper-parameters collected in $\eta = [ \lambda,\ \beta]$. The length $n$ of the estimated impulse responses has been set to 80.

\subsection{Monte-Carlo study on BIBO stable time invariant systems}
For each of the 200 Monte-Carlo runs we consider in our study we have generated a random SISO discrete-time system through the Matlab routine \texttt{drmodel.m}. The system orders have been randomly chosen in the range $[5,10]$, while the systems poles are all inside a circle of radius 0.95. The input signal is a unit variance band-limited Gaussian signal with normalized band $[0,0.8]$. A zero mean white Gaussian noise, with variance adjusted so that the Signal to Noise Ratio (SNR) is always equal to 5, has been added to the output data. For each Monte-Carlo run the total number of available data is  $N=5000$, while the length of the on-line upcoming datasets $\mathcal{D}_k$ has been chosen to be $N_k = 10$; furthermore, the on-line Algorithm \ref{alg:on_line_setup} is initialized by computing the OPT procedure on the first 100 data.

In the interest of reducing the computational time of the on-line updates we propose two versions of BFGS, SGP, BB, EM: the first one updates both the hyper-parameters in $\eta$ whenever a new dataset $\mathcal{D}_k$ becomes available, while the second one updates only the scaling factor $\lambda$, retaining $\beta$ fixed to its initial value. It is clear that the latter case allows a faster computation, at the expense of a less precise impulse response estimator. In addition, two cases of the EM version which only updates $\lambda$ are considered: $EM2$, where the update corresponds to \eqref{equ:em_lambda_update} and $EM1$, where $\widehat{\lambda}^{(k+1)}=\frac{1}{n}\widehat{\mathbf{h}}^{(k)^\top}K_{\hat{\beta}}^{-1}\widehat{\mathbf{h}}^{(k)}$, which is the current approximation of the asymptotically optimal value. The aim is to show a comparison between the asymptotic theory and the EM update, see e.g. \cite{GB-AYA-HH-GP:14}; notice that the second term of \eqref{equ:em_lambda_update} tends to zero when the number of data tends to infinity.

As a first comparison, we evaluate the adherence of the impulse response estimate to the true one. Thus, for each estimated system and for each procedure we compute the impulse response fit:
\begin{equation}
\label{eq:fit_imp_resp}
\mathcal{F}(\widehat{\mathbf{h}})= 100 \cdot \Big(1- \frac{\Vert \mathbf{h} - \widehat{\mathbf{h}} \Vert_2}{\Vert \mathbf{h} \Vert_2}\Big)
\end{equation}
where $\mathbf{h}, \, \widehat{\mathbf{h}}$ are the true and the estimated impulse responses of the considered system, respectively.

\begin{figure}[!h]
\begin{center}
\includegraphics[scale=0.45]{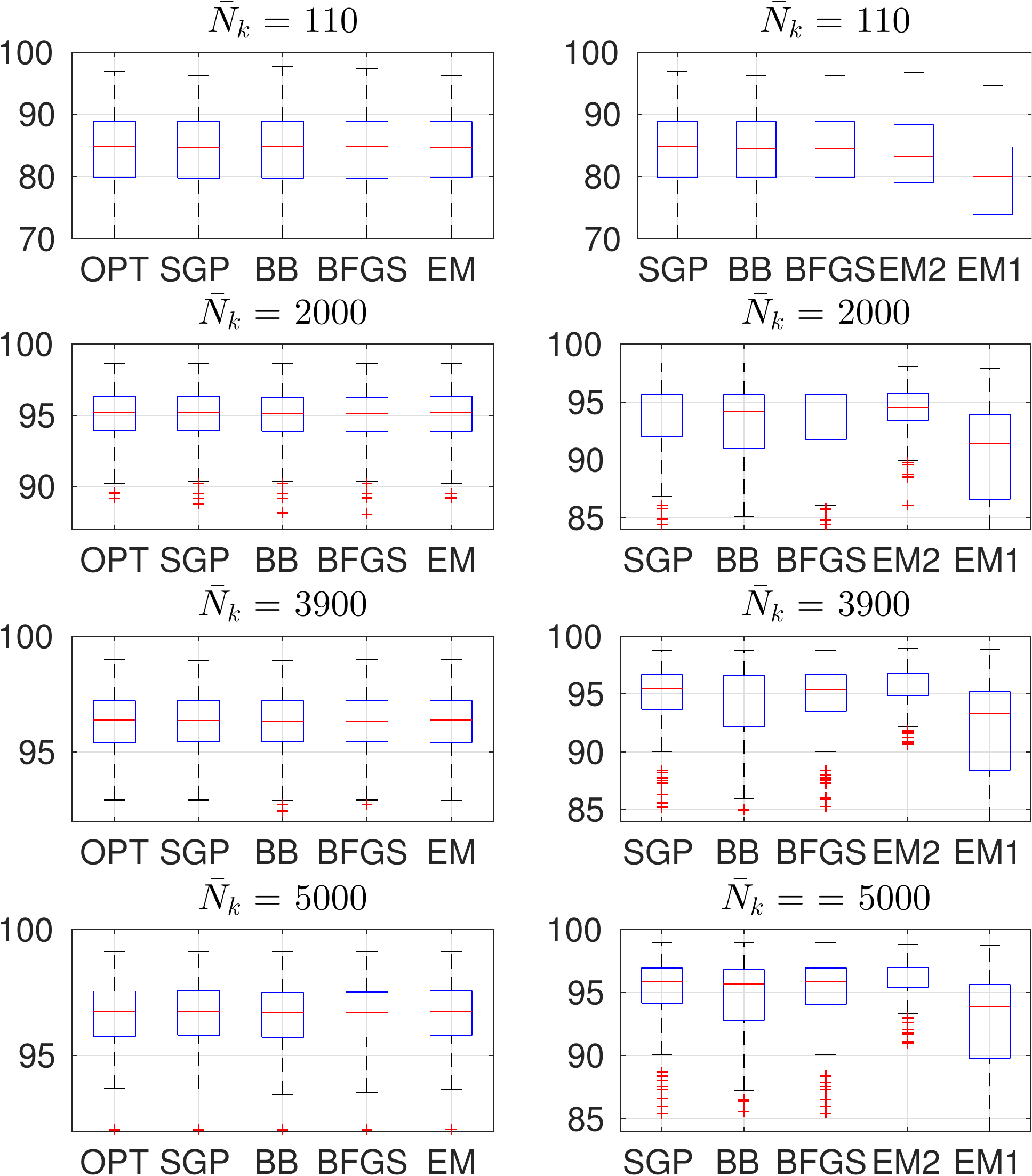} 
\caption{Monte Carlo results. \textit{Left:} Boxplots of the impulse response fit obtained updating both hyper-parameters in $\eta$. \textit{Right}: Boxplots of the impulse response fit obtained updating only $\lambda$.}
\label{fig:fit_boxplot}
\end{center}
\end{figure}

Figure \ref{fig:fit_boxplot} shows the impulse response fits \eqref{eq:fit_imp_resp} achieved in the Monte-Carlo simulations we considered along with the increase of the number of observed data. OPT procedure is compared with the 1-STEP SGP, BB, BFGS and EM. On the left hand side the results obtained optimizing both the hyper-parameters in $\eta$ are reported, while  the results on the right hand side are obtained by updating only $\lambda$. 

\vspace{-2mm}

All the 1-STEP procedures which update both the hyper-parameters perform remarkably well, with the fit index being almost equivalent to the one obtained with the OPT procedure. This suggests that the full optimization of problem \eqref{equ:ml_max} does not bring any particular advantage in terms of fit in the on-line setting. Notice that we are taking a sort of worst case approximation since we are stopping the optimization algorithm after only 1 step: some more evolute techniques could be considered (e.g. an early stopping criterion \cite{YY-RL-CP:07}). The 1-STEP updates optimizing only $\lambda$, after a transient period, perform comparably (but slightly worse) to the other techniques; the only exception is represented by EM1 which achieves inferior fits, but we expect that also this update reaches the same performances when the number of data tends to infinity.

\begin{figure}[!h]
\begin{center}
\includegraphics[scale=0.45]{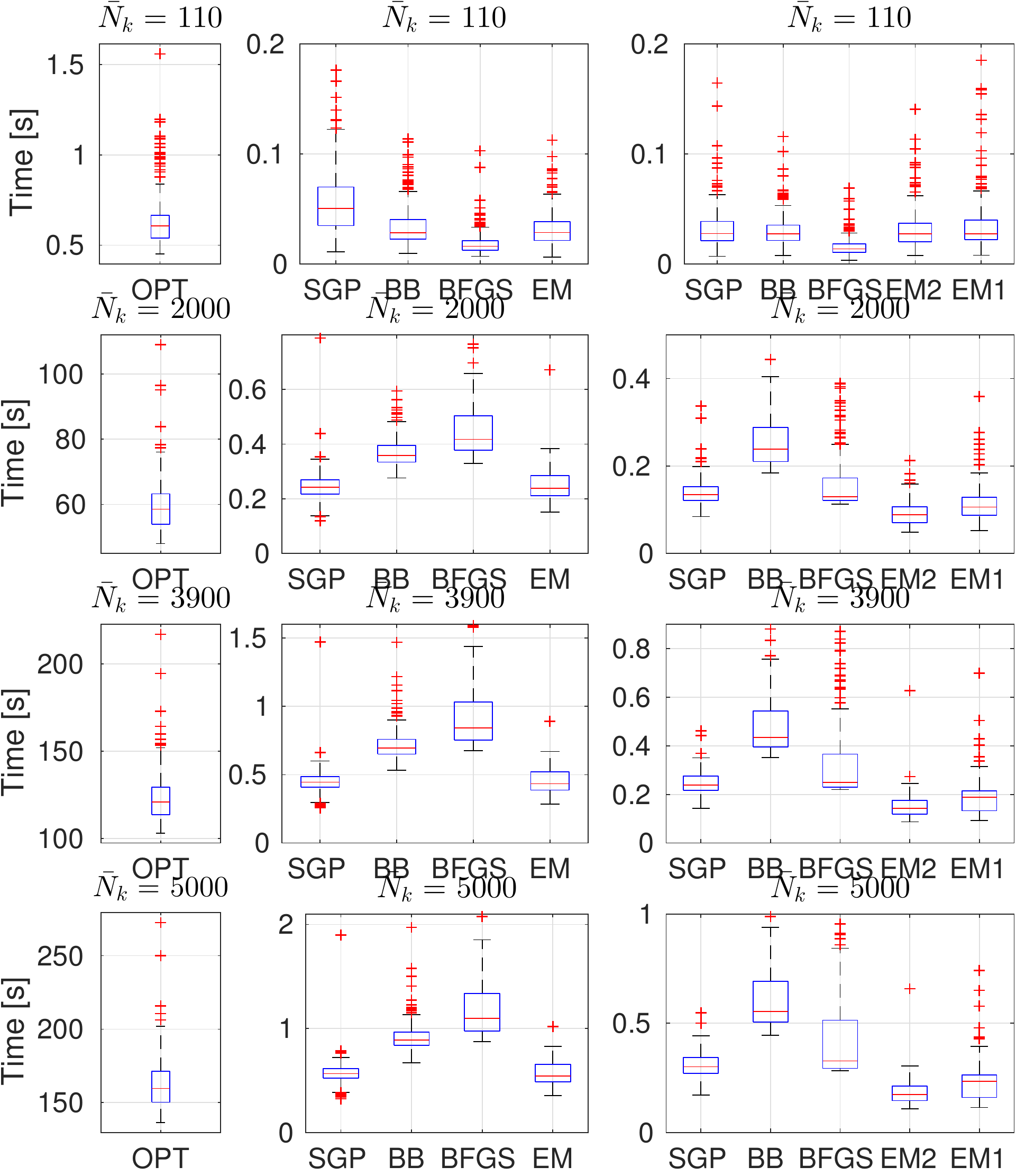} 
\caption{Monte Carlo results. Boxplots of the cumulative computational time. Each row of plots corresponds to the situation after $N_k$ data are viewed. \textit{Left:} OPT procedure. \textit{Mid:} 1-STEP optimization of both hyper-parameters. \textit{Right:} 1-STEP optimization only of $\lambda$ ($\beta$ is fixed).}
\label{fig:cumulative_time}
\end{center}
\end{figure}

The second comparison is done in terms of cumulative computational time of the procedures, see Figure \ref{fig:cumulative_time}.

\begin{figure}[!h]
\begin{center}
\includegraphics[scale=0.45]{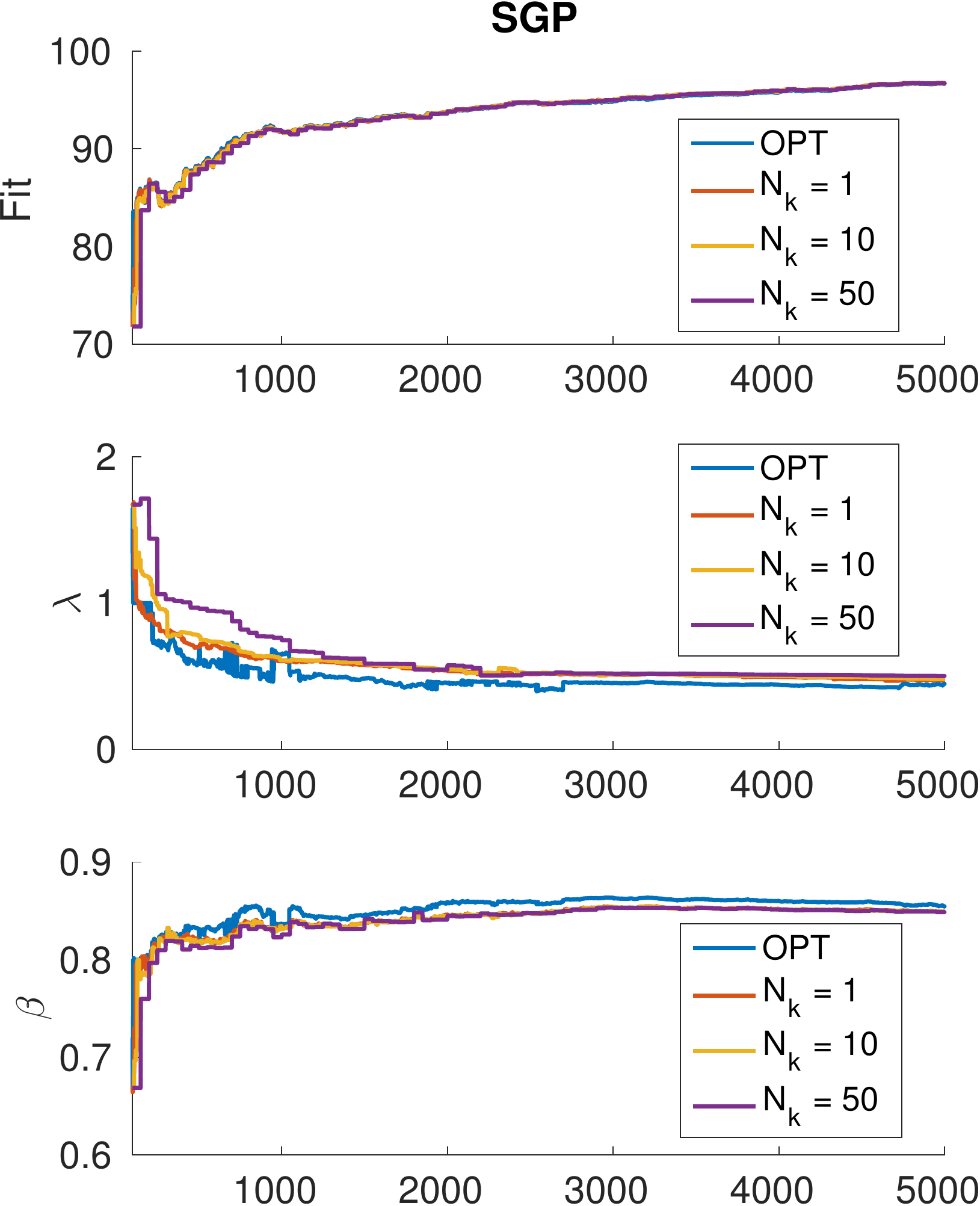} 
\caption{Comparison of OPT and 1-STEP update with different length $N_k$ of the dataset $\mathcal{D}_k$ in the on-line identification of one system.}
\label{fig:sing_sys}
\end{center}
\end{figure}

The OPT procedure, as expected, is much slower than the 1-STEP procedures. This could suggest that the 1-STEP procedures we consider appear to be excellent candidates for real-time applications. Indeed, these techniques perform comparably in terms of fit w.r.t. the OPT procedure, but demanding a computational time which is two or three order faster; furthermore the difference in terms of computational time diverges in favour of the 1-STEP procedure with the increase of the number of data seen. Among the 1-STEP procedures SGP and EM provide the fastest updates: this is surprisingly positive for the EM update since only $\lambda$ has a closed form update, while $\beta$ is the solution of a maximization problem; indeed, in the right hand side of Figure \ref{fig:cumulative_time}, where only $\lambda$ is updated, EM1 and EM2 outperform SGP. The update BB is a particular case of SGP, where $D^{(k)}=I$ (see Section \ref{Sec:1stepMLmax}), but it is significantly slower: this is due to the computation of the projection step \ref{alg_step:proj} in Algorithm \ref{alg:1step_grad}.  In the right hand side of Figure \ref{fig:cumulative_time} we can see the advantage of updating only $\lambda$: the cumulative computational time is inferior.
%
Finally, in Figure \ref{fig:sing_sys} we show the evolution of the fit and of the hyper-parameters estimates when new datasets arrive for a single system. In this experiment,  we compare cases with different lengths of the datasets $\mathcal{D}_k$, i.e. $N_k = 1, 10, 50$. We can notice how the results do not differ significantly among the considered values of $N_k$.
%


\section{Conclusions and Future Work}\label{sec:conclusion}
We have considered the application of Bayesian identification techniques in an on-line setting. In order to meet real-time requirements, reducing the computational time required to update the impulse response estimate becomes essential. In a Bayesian estimation procedure, the most demanding step in terms of computational complexity is the Marginal Likelihood optimization required to determine the hyper-parameters estimate. In this work we have considered different iterative procedures that are typically used to solve the Marginal Likelihood maximization problem. Moreover, in order to address the real-time requirements, we proposed to update the hyper-parameters by only performing one iteration of the above-mentioned techniques.
The experimental results we have shown seem very promising. \\
Future work will include adaptations to track (slowly) time varying dynamics as well as further simplifications on the computational aspects, which have not been yet fully optimized in this preliminary study.  

 

%
%
%
%
%

\bibliographystyle{IEEEtran}
\bibliography{References}

\end{document}